\documentclass[useAMS,usenatbib]{mn2e}
\usepackage{graphicx}
\def\apj{ApJ}
\def\apjs{ApJS}

\def\aap{A\&A}

\def\aj{AJ}
\def\mnras{MNRAS}

\def\pasp{PASP}

\def\araa{Ann.Rev.Astron.Astrophys.}

\title[Universality of mass-metallicity relation]
{On the universality of luminosity-metallicity and mass-metallicity 
relations for compact star-forming galaxies at redshifts 0 $<$ $z$ $<$ 3}
\author[Y. I. Izotov et al.]{Y. I.\ Izotov$^{1,2,3}$,
N. G.\ Guseva$^{1,2}$, K. J.\ Fricke$^{2,4}$ and C.\ Henkel$^{2,5}$\\
                $^{1}$Main Astronomical Observatory,
                     Ukrainian National Academy of Sciences,
                     Zabolotnoho 27, Kyiv 03680,  Ukraine\\
                $^{2}$Max-Planck-Institut f\"ur Radioastronomie, 
                     Auf dem H\"ugel 
                     69, 53121 Bonn, Germany\\
                $^{3}$LUTH, Observatoire de Paris, CNRS, 
                     Universite Paris Diderot,
                     Place Jules Janssen 92190 Meudon, France\\
                $^{4}$Institut f\"ur Astrophysik, 
                     G\"ottingen Universit\"at, Friedrich-Hund-Platz 1, 
                     37077 G\"ottingen, Germany\\
                $^{5}$Astronomy Department, King Abdulaziz University, 
                     P.O. Box 80203, Jeddah 21589, Saudi Arabia\\
}
\begin{document}

\date{Accepted 1988 December 15. Received 1988 December 14; in original form 1988 October 11}

\pagerange{\pageref{firstpage}--\pageref{lastpage}} \pubyear{2012}

\maketitle

\label{firstpage}

\begin{abstract}
We study relations between global characteristics of low-redshift 
(0 $<$ $z$ $<$ 1) compact star-forming galaxies, including 
absolute optical magnitudes,
H$\beta$ emission-line luminosities (or equivalently star-formation rates),
stellar masses, and oxygen abundances. The sample consists of 5182 galaxies
with high-excitation H {\sc ii} regions selected from the SDSS DR7 and
SDSS/BOSS DR10 surveys adopting a criterion 
[O~{\sc iii}]$\lambda$4959/H$\beta$~$\geq$~1. 
These data were combined with the corresponding
data for high-redshift (2~$\la$~$z$~$\la$~3) star-forming galaxies.
We find that
in all diagrams low-$z$ and high-$z$ star-forming
galaxies are closely related indicating a very weak dependence 
of metallicity on stellar mass, redshift, and star-formation rate.
This finding argues in favour of the universal character of the global
relations for compact star-forming
galaxies with high-excitation H~{\sc ii} regions over redshifts 0~$<$~$z$~$<$~3.
\end{abstract}


\section{Introduction}\label{intro}

Luminosities, stellar masses, and
metallicities are among the most important properties of galaxies,
which are needed for a better understanding of their formation and evolution.
The H$\beta$ emission-line luminosity allows for a direct measure of
the rate of gas transformation into stars, while stellar mass
and metallicity are indicators of the galaxy's past evolution.
Thus, relations between these parameters and their
variations with redshift may help to analyse the origin and
evolution of galaxy populations on cosmological time scales.

In this respect, the study of star-forming galaxies plays a
particularly important role. The brightness of these galaxies is enhanced 
by the ongoing star-formation, allowing us to observe them at
larger distances. Furthermore, stronger emission lines in spectra
of these galaxies better constrain their metallicity, one of
the key galaxy global parameters.

The Sloan Digital Sky Survey (SDSS) imaging and spectroscopy
\citep{Y00} opened a unique opportunity
to produce the luminosity-metallicity and mass-metallicity relations
for large samples of low-$z$ normal and star-forming galaxies.
\citet{T04} obtained a mass-metallicity relation for $\sim$53000 
SDSS star-forming galaxies at
$z$ $\sim$ 0.1. They found that the relation is relatively steep for
low-mass galaxies with stellar masses $M_*$ $<$ 10$^{10.5}$$M_\odot$, but 
flattens for higher-mass galaxies. The mass-metallicity relation
by \citet{T04} is often used as a benchmark for comparisons with
similar relations obtained for low-$z$ and high-$z$ galaxies
\citep[e.g. ][]{L06,M08,A10,M10,M14,S14,Z11,Z12,Z13,Z14a,Z14b}. 
However, we note that \citet{T04} selected all star-forming galaxies,
including galaxies with low-excitation H~{\sc ii} regions, 
while only galaxies with high-excitation H~{\sc ii} regions
are present in high-$z$ samples. They also used calibrations by \citet{CL01}
for the metallicity determination, which were not used for high-$z$ galaxies.
This implies that differences between
low-$z$ and high-$z$ galaxies can be introduced, at least in part,
by different selection criteria and different methods used for the
determination of global galaxy parameters such as metallicities and stellar 
masses.

\citet{M08} found a strong evolution of
the stellar mass-metallicity relation with redshift in the sense
that high-$z$ ($z$ $\sim$ 3.5) galaxies
are more metal-poor compared to low-$z$ galaxies for the same stellar
mass. This effect is more pronounced for low-mass galaxies.
\citet{M10} considered a more general relation between stellar mass $M_*$,
metallicity, and star-formation rate (SFR). They found that high-$z$
galaxies are characterised by progressively higher SFRs. Introducing
a parameter $\mu$=log$M_*/M_\odot$ - $\alpha$logSFR($M_\odot$yr$^{-1}$), \citet{M10}
showed that, independent of redshift, star-forming galaxies
have a uniform $\mu$ -- metallicity distribution if $\alpha$ = 0.32.

On the other hand, there is some evidence that properties of 
nearby and high-$z$ star-forming galaxies are similar with respect to SFR 
and metallicity.
\citet{H05} have
identified nearby ($z$ $<$ 0.3) ultraviolet-luminous galaxies (UVLGs) selected
from the {\sl Galaxy Evolution Explorer} ({\sl GALEX}). These compact UVLGs 
were eventually called Lyman-break analogues (LBAs). They resemble LBGs in 
several respects, implying that mass-metallicity relations for low-$z$ and
high-$z$ galaxies should be similar.  In particular, their metallicities are 
subsolar, and their SFRs of 
$\sim$ 4 -- 25 $M_\odot$ yr$^{-1}$ are overlapping with those for LBGs.

Furthermore, \citet{C09} selected a sample of compact strongly 
star-forming galaxies from the SDSS, which 
are also similar to LBGs because of their low metallicities and high SFRs. 
\citet{I11a} extracted a sample of star-forming luminous compact
galaxies (LCGs) with hydrogen H$\beta$ luminosities 
$L$(H$\beta$) $\geq$ 3$\times$10$^{40}$ erg s$^{-1}$ and H$\beta$ equivalent
widths EW(H$\beta$) $\geq$ 50\AA\ from SDSS spectroscopic data.
These galaxies have properties 
similar to ``green pea'' galaxies \citep{C09} but are distributed over a wider
range of redshifts $z$ $\sim$ 0.02 - 0.63. The SFRs of LCGs are high at
$\sim$ 0.7 -- 60 $M_\odot$ yr$^{-1}$ and overlap with those of LBGs.

\citet[see also \citet{G09} and \citet{Zh10}]{I11a} showed that LBGs, LCGs, 
luminous metal-poor
star-forming galaxies \citep{Hoyos05}, extremely metal-poor emission-line 
galaxies at $z$ $<$ 1 \citep{Kakazu07}, 
and low-redshift blue compact
dwarf (BCD) galaxies with strong star-formation activity 
obey a common luminosity-metallicity relation. Finally, \citet{I14a}
considered luminosity-metallicity and mass-metallicity relations for SDSS DR7
star-forming galaxies, which are much flatter than the \citet{T04} relations.
They also found that low-$z$ galaxies with high SFRs do not deviate from
the relation established for galaxies with lower SFRs.

The above discussion implies some inconsistency between results obtained
in different papers and raises the question: are luminosity- and 
mass-metallicity relations between global parameters of low-$z$ and high-$z$
star-forming galaxies different? What is the impact of
different selection criteria and methods used for the determination
of the galaxy metallicities?
In this paper we attempt to study similarities
and differences of the relations for star-forming galaxies 
in a redshift range 0 $<$ $z$ $<$ 3. Special care has been taken to
use strong-line methods to derive the metallicity, which were
calibrated using the $T_{\rm e}$-method, based on the electron
temperature determination from the 
[O~{\sc iii}]($\lambda$4959+$\lambda$5007)/$\lambda$4363 flux ratio 
\citep[e.g. ][]{I11a}, and thus give consistent gas-phase
metallicities. We also selected only compact star-forming galaxies with
high-excitation H~{\sc ii} regions, for which both methods can be applied.
Furthermore, excitation conditions in those H~{\sc ii} regions are similar
to those in high-$z$ galaxies.

In Sect. 2 we discuss the sample. The technique used for the determination
of luminosities and stellar masses is described in Sect. 3. Different methods
of the oxygen abundance determination are discussed in Sect. 4. Relations
between different global parameters for our SDSS sample and comparisons
with respective relations for high-$z$ galaxies with high-excitation
H~{\sc ii} regions are presented in Sect. 5.
The main results of the paper are described in Sect. 6.

\begin{figure}
\includegraphics[angle=-90,width=0.99\linewidth]{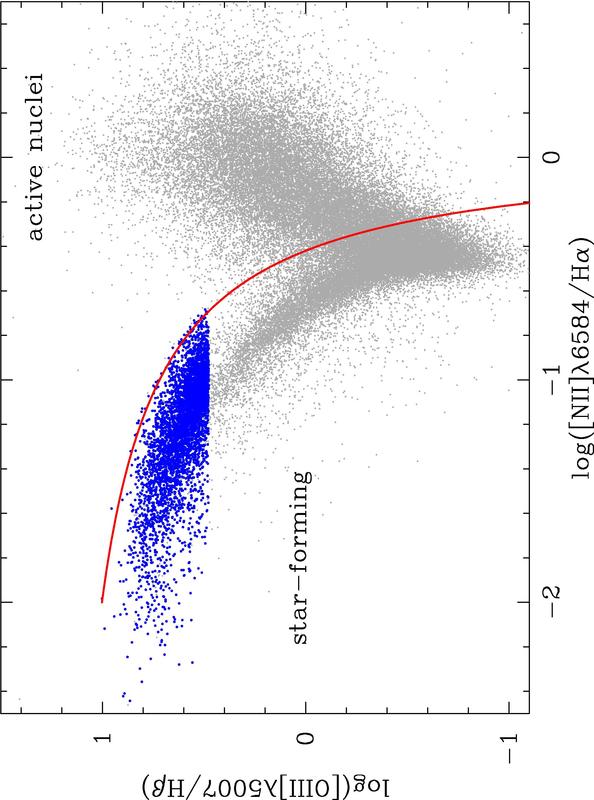}
\caption{ The Baldwin-Phillips-Terlevich (BPT) 
diagnostic diagram \citep{B81}.
Selected SDSS compact star-forming galaxies with 
[O~{\sc iii}]$\lambda$4959/H$\beta$ $\geq$ 1.0, 
corresponding to log~[O~{\sc iii}]$\lambda$5007/H$\beta$ $\ga$ 0.5,
and detected [N~{\sc ii}]$\lambda$6584 line are shown by blue filled
circles. Also plotted are all emission-line galaxies from SDSS DR7
(cloud of grey dots). The red solid line from \citet{K03}
separates star-forming galaxies from active galactic nuclei.}
\label{fig1}
\end{figure}

\begin{figure}
\begin{center}
\hspace{0.5cm}\includegraphics[angle=0,width=0.85\linewidth]{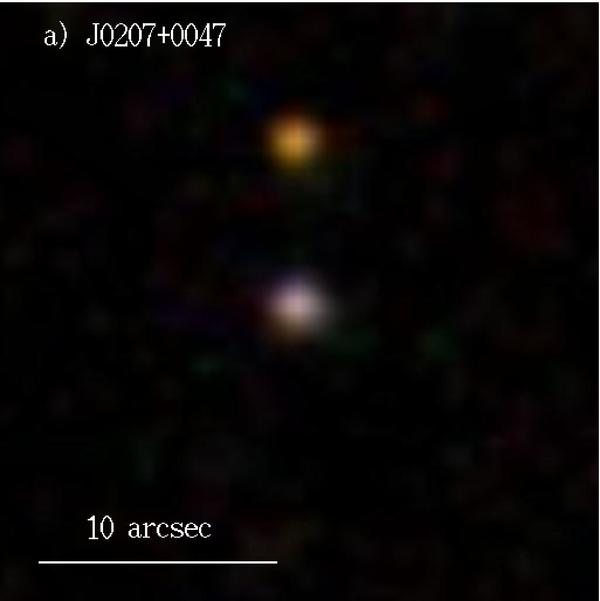}
\end{center}
\vspace{0.2cm}
\begin{center}
\includegraphics[angle=-90,width=0.99\linewidth]{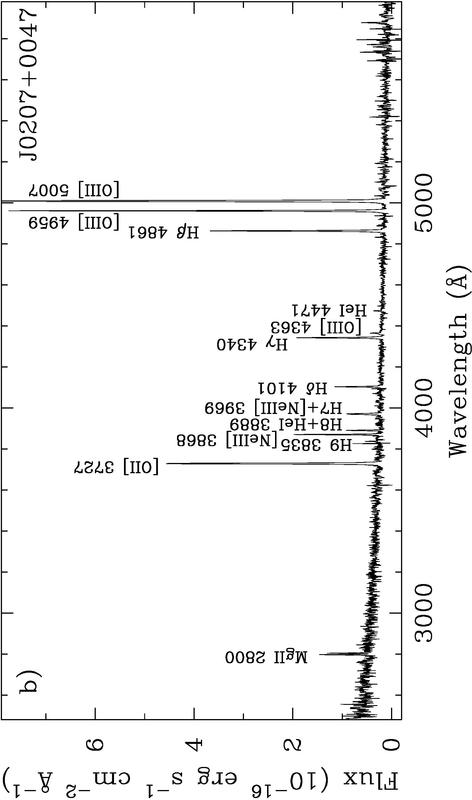}
\end{center}
\caption{ (a) The 25\arcsec$\times$25\arcsec\ SDSS image centered on the
compact star-forming galaxy J0207+0047. (b) The rest-frame SDSS spectrum of 
J0207+0047. The galaxy is located at a redshift of $z$ = 0.54.}
\label{fig2}
\end{figure}

\begin{figure}
\includegraphics[angle=-90,width=0.99\linewidth]{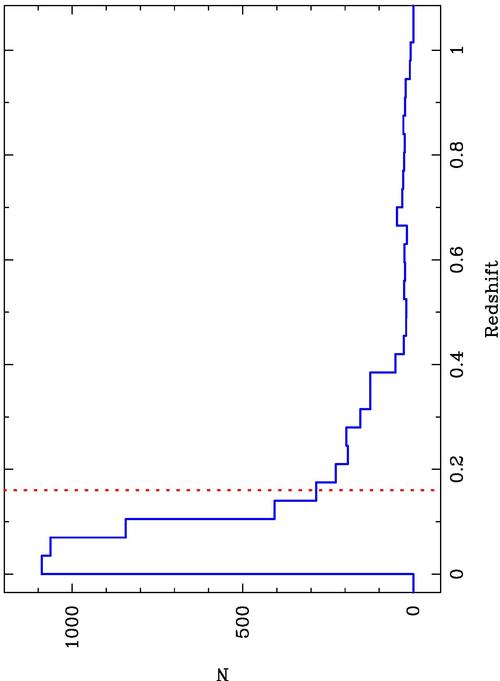}
\caption{ Distribution over redshift $z$ of the SDSS compact star-forming 
galaxies from the entire sample. 
The dotted vertical line indicates the average redshift of 0.16.}
\label{fig3}
\end{figure}

\section{The sample \label{sel}}

   The sample of compact star-forming galaxies was selected from the 
spectroscopic data base of the SDSS DR7 \citep{A09} and SDSS/BOSS DR10
\citep{A14,D13,S13}. The criteria were:
1) only galaxies with angular diameters $\la$ 6\arcsec\ were selected
to minimize the aperture corrections of the derived luminosities
and stellar masses. As a measure of the angular size we used the
Petrosian radius $R_{50}$ with a 50\% flux of the galaxy
inside it in the SDSS $r$ band. These 
data are available in the SDSS data base. The oxygen abundance is then 
characterizing the entire galaxy because of its compactness; 
2) spiral galaxies are excluded. The oxygen abundances in
individual H {\sc ii} regions are not representative characteristics for
an entire spiral galaxy because of the abundance
gradients in those galaxies;
3) galaxies with emission-line ratios 
[O~{\sc iii}]$\lambda$4959/H$\beta$ $<$ 1.0 were excluded to ensure
that only galaxies with high-excitation H {\sc ii} regions are included,
for which the determination of extinction and chemical composition 
is more reliable. We note that most of  high-redshift emission-line 
galaxies satisfy the criterion [O~{\sc iii}]$\lambda$4959/H$\beta$ $\geq$ 1.0,
and hence have similar excitation conditions in their H~{\sc ii} regions; 
4) galaxies with AGN activity are excluded. We used
obvious AGN spectral characteristics: broad emission lines (QSOs and 
Sy1 galaxies), the presence of strong high-ionisation 
[Ne~{\sc v}] $\lambda$3426 and 
He~{\sc ii} $\lambda$4686 emission lines in high-excitation spectra of
Sy2 galaxies. Strong high-ionisation nebular [O~{\sc iii}] emission lines are 
simultaneously observed with strong low-ionisation nebular [O~{\sc i}], 
[O~{\sc ii}], [N~{\sc ii}], and [S~{\sc ii}] emission lines in many 
high-excitation Sy2 spectra. 
LINERs are excluded by a condition 
[O~{\sc iii}]$\lambda$4959/H$\beta$ $<$ 1.0. The reliability of our selection 
is checked with the diagnostic diagram, proposed by \citet{B81} (BPT diagram, 
Fig. \ref{fig1}). All galaxies from our sample are located in the region 
of star-forming galaxies (blue dots in Fig. \ref{fig1}).

Applying these criteria, we selected 5182 galaxies in the redshift range
0 $<$ $z$ $<$ 1.
The [O~{\sc iii}] $\lambda$4363 emission line was detected at a level
higher than 1$\sigma$ in 3607 galaxies from our sample, allowing us to
obtain oxygen abundances by the
$T_{\rm e}$-method. Hereafter this subsample will be called 
``$T_{\rm e}$ sample''. We apply strong-emission-line (SEL) methods 
to derive oxygen abundances for the remaining subsample of 1575 galaxies.

The SDSS image of one compact galaxy from our sample, J0207+0047, is 
shown in Fig. \ref{fig2}a. This galaxy is located at the redshift 
$z$ = 0.54 and is very compact and is almost unresolved. Its angular diameter
is only 2$R_{50}$ = 1.4\arcsec, corresponding to a linear diameter of
$\sim$ 30 kpc. 
However, since this galaxy is unresolved, its true linear
diameter can be much smaller. This is a common property for distant galaxies
from our sample with $z$ $\ga$ 0.2 -- 0.3. {\sl HST} images of a few 
compact star-forming galaxies at $z$ $\sim$ 0.1 -- 0.35 reveal that all of
them have linear diameters less than 3 kpc \citep[][Izotov et al. 2015, in
preparation]{C09,JO14}. The spectrum of J0207+0047 
(Fig. \ref{fig2}b) is characterised by strong narrow emission lines 
and blue continuum. Spectra of all our galaxies have similar characteristics.
They are very different from the spectra of ``normal'' galaxies
dominated by numerous absorption lines and much redder continua. 
The spectrum in Fig. \ref{fig2}b is also very different from 
typical spectra of star-forming galaxies with weak emission lines, which
were selected by \citet{T04}. 

The redshift distribution of our entire sample with the average redshift
of 0.16 (dotted vertical line) is shown in Fig.~\ref{fig3}.

Spectroscopic data were supplemented with photometric data in five SDSS bands 
$u$, $g$, $r$, $i$, and $z$ for the entire galaxy
and within the spectroscopic aperture,
allowing aperture corrections of the observed fluxes and the determination
of absolute galaxy magnitudes.

\section{The determination of galaxy global parameters}

\subsection{Parameters of emission lines and dust extinction}

We measured emission-line fluxes and equivalent widths using the 
IRAF\footnote {IRAF is the Image 
Reduction and Analysis Facility distributed by the National Optical Astronomy 
Observatory, which is operated by the Association of Universities for Research 
in Astronomy (AURA) under cooperative agreement with the National Science 
Foundation (NSF).} SPLOT routine. The line-flux errors 
include statistical errors in addition to errors introduced by
the standard-star absolute flux calibration, which we set to 1\% of the
line fluxes. These errors will be later propagated into the calculation
of abundance errors.
Using the observed decrement of several
hydrogen Balmer emission lines we corrected the line fluxes relative to the 
H$\beta$ flux for two effects: (1) reddening adopting
the extinction curve of \citet{C89} and 
(2) underlying hydrogen stellar absorption that is derived simultaneously by an
iterative procedure as described in \citet{ITL94}.
The extinction coefficients 
are defined as $C$(H$\beta$)~=~1.47$E(B-V)$,
where $E(B-V)$~=~$A(V)$/3.2 \citep{A84}. 

The spectroscopic data were corrected for the aperture using the relation 
2.5$^{r({\rm app})-r}$, where $r$ and $r$(app) are the 
SDSS $r$-band total magnitude
and the magnitude within the round spectroscopic aperture, respectively.
The diameters of spectroscopic apertures are 3\arcsec\ and 2\arcsec\ for the 
SDSS DR7 and SDSS/BOSS spectra, respectively. Selected galaxies are
compact. Therefore the aperture corrections of fluxes are 
small, rarely exceeding the value of 2.

\subsection{The absolute $g$-band magnitudes and H$\beta$ luminosities}

The extinction-corrected absolute $g$-band magnitudes $M_g$ and 
extinction- and aperture-corrected H$\beta$ luminosities $L$(H$\beta$) were obtained 
respectively from the SDSS extinction-corrected $g$ magnitude for the entire 
galaxy and extinction- and aperture-corrected H$\beta$ emission-line flux.
The distance is derived from the redshift. For distance determination 
we used the relation $D$ = $f$($z$,$H_0$,$\Omega_{\rm M}$,$\Omega_\Lambda$)
from \citet{R67}, where
the Hubble constant $H_0$ = 67.3 km s$^{-1}$ Mpc$^{-1}$ and cosmological
parameters, $\Omega_{\rm M}$ = 0.273 and $\Omega_\Lambda$ = 0.682, 
were obtained from
the {\sl Planck} mission data \citep{Ade14}. The equivalent widths EW(H$\beta$)
were correspondingly reduced to the rest frame.


\begin{figure*}
\hbox{
\includegraphics[angle=-90,width=0.48\linewidth]{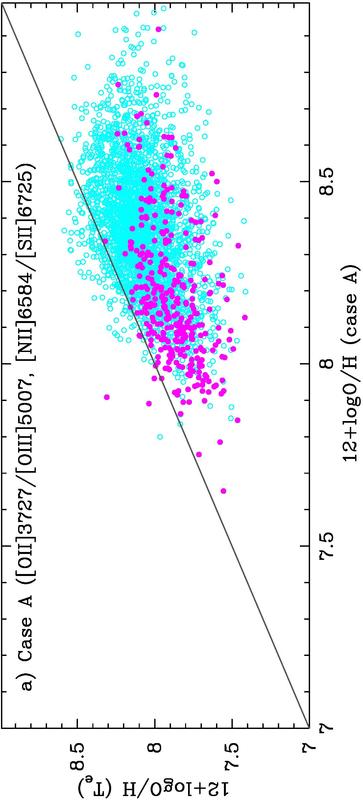}
\hspace{0.2cm}\includegraphics[angle=-90,width=0.48\linewidth]{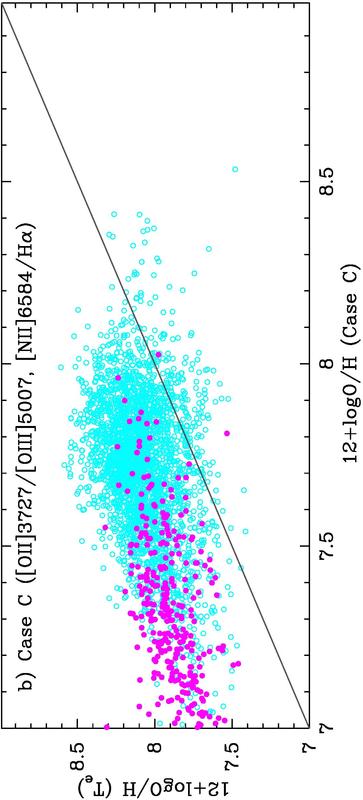}
}
\vspace{0.2cm}
\hbox{
\includegraphics[angle=-90,width=0.48\linewidth]{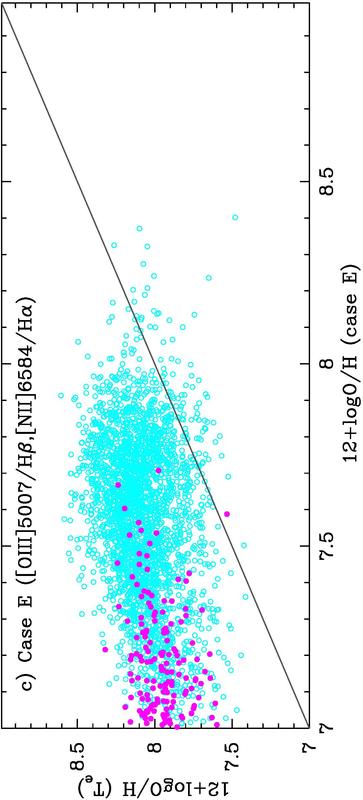}
\hspace{0.2cm}\includegraphics[angle=-90,width=0.48\linewidth]{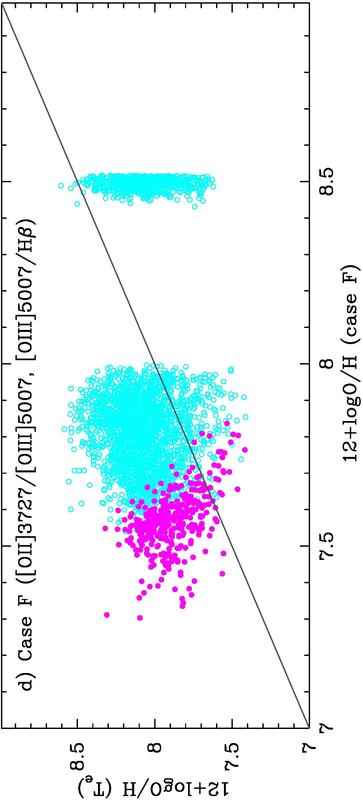}
}
\caption{A comparison of oxygen abundances 12 + logO/H, derived by the
$T_{\rm e}$-method (this paper) and different strong-line methods from 
\citet{CL01} for our $T_{\rm e}$ sample. 
In each panel, the emission-line ratios used by strong-line 
methods are given in parentheses. Galaxies with 
$O_{32}$ =[O~{\sc iii}]$\lambda$5007/[O~{\sc ii}]$\lambda$3727 
$\geq$ 4.5 and $<$ 4.5 are given by magenta filled circles and cyan open 
circles, respectively. Solid lines show the one-to-one relations.}
\label{fig4}
\end{figure*}

\subsection{Spectral energy distribution (SED) and the galaxy's stellar masses}

The stellar mass is one of the most important global galaxy
characteristics. For its determination we follow the prescriptions
described by \citet{G06,G07} and \citet{I11a,I14a,I14b}.

The method is based on fitting a series of model SEDs to the observed one 
and finding the best fit. It consists of the following. The fit was performed 
for each SDSS spectrum over the entire
observed spectral range of $\lambda$$\lambda$3900--9200\AA\
for SDSS DR7 galaxies and of $\lambda$$\lambda$3600--10300\AA\ for
SDSS/BOSS galaxies.
As each SED is the sum of both stellar and ionised gas emission,
its shape depends on the relative contribution of these two
components. In galaxies with high EW(H$\beta$)~$>$~50\AA, 
the ionised gas continuum is strong and is
subtracted before fitting the stellar SED.
\citet{I11a} noted the importance of gaseous continuum subtraction,
otherwise stellar masses of galaxies with high EW(H$\beta$)'s
in their spectra would be overestimated by a factor of $\sim$ 3 or more.
Hence, for the spectra of galaxies with EW(H$\beta$)~$>$~50\AA,
SED fitting and subtraction of the ionised gas emission
in the continuum and emission lines are mandatory.
Photometric data are not sufficient in these cases.

We carried out a series of Monte Carlo simulations to
reproduce the SED of each galaxy in our sample. To derive
the stellar SED, we use a grid of instantaneous burst SEDs in a wide range 
of ages from 0.5 Myr to 15 Gyr calculated with package PEGASE.2 
\citep{FR97}. We adopted a stellar initial mass 
function with a Salpeter slope, an upper mass limit of 100~$M_\odot$, and a 
lower mass limit of 0.1~$M_\odot$. Then the SED with any star-formation 
history can be obtained by integrating the instantaneous burst SEDs over
time with a specified time-varying star-formation rate. 

The SED of the gaseous continuum was
taken from \citet{A84}.
It included hydrogen and helium free-bound, free-free, and two-photon emission.

The star-formation history in each galaxy is approximated assuming a recent 
short burst with age $t_{\rm y}$ $<$ 10 Myr, which accounts 
for the young stellar population, and a prior continuous star formation 
for the older stars during the time interval between $t_i$ and $t_f$
($t_f < t_i$ and zero age is now).
The contribution of each stellar population to the SED was parameterized by the
ratio $b$=$M_{\rm y}$/$M_{\rm o}$, where $M_{\rm y}$ and $M_{\rm o}$ 
are respectively the masses of the young and old stellar populations.
Then the total stellar mass is $M_*$= $M_{\rm y}$ + $M_{\rm o}$.

For each galaxy, we calculated 
10$^4$ Monte Carlo models by randomly varying $t_{\rm y}$, $t_i$,
$t_f$, and $b$. The best
modelled SED was found from $\chi ^2$ minimization of the deviation between the
modelled and the observed continuum in five wavelength ranges, which are free 
of the emission lines and residuals of the night-sky lines. 
Typical stellar mass uncertainties for our sample galaxies are 
of $\sim$ 0.1 -- 0.2 dex.


\begin{figure*}
\hbox{
\includegraphics[angle=-90,width=0.48\linewidth]{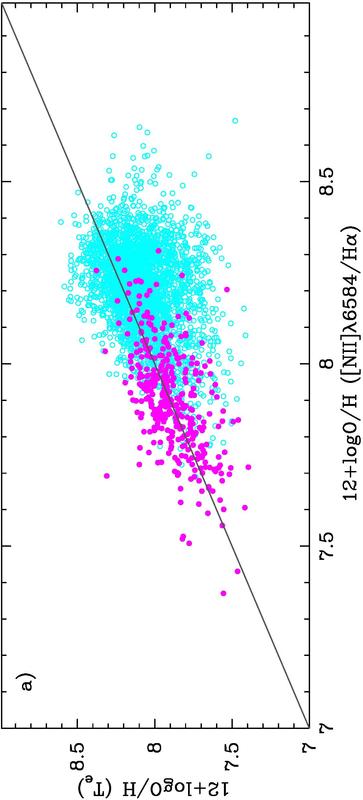}
\hspace{0.2cm}\includegraphics[angle=-90,width=0.48\linewidth]{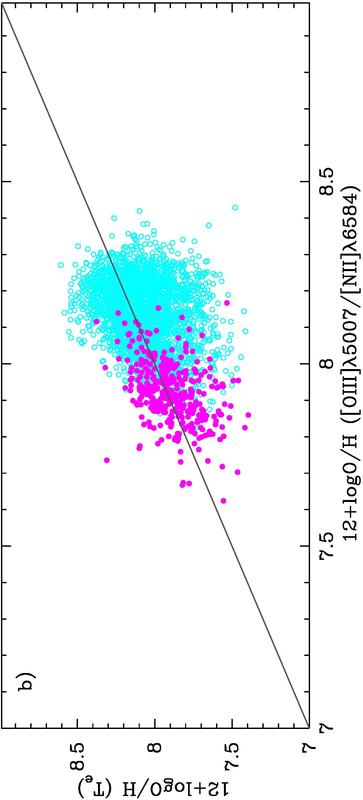}
}
\caption{A comparison of oxygen abundances 12 + logO/H derived by the
$T_{\rm e}$-method (this paper) 
and strong-emission-line (SEL) methods proposed by \citet{PP04}, which
utilize the (a) [N~{\sc ii}]$\lambda$6584/H$\alpha$ and 
(b) [O~{\sc iii}]$\lambda$5007/[N~{\sc ii}]$\lambda$6584 line-intensity ratios.
Shown are the same galaxies by the same symbols as in Fig.~\ref{fig4}. 
Solid lines denote the one-to-one correlations.
}
\label{fig5}
\end{figure*}


\begin{figure*}
\hbox{
\includegraphics[angle=-90,width=0.48\linewidth]{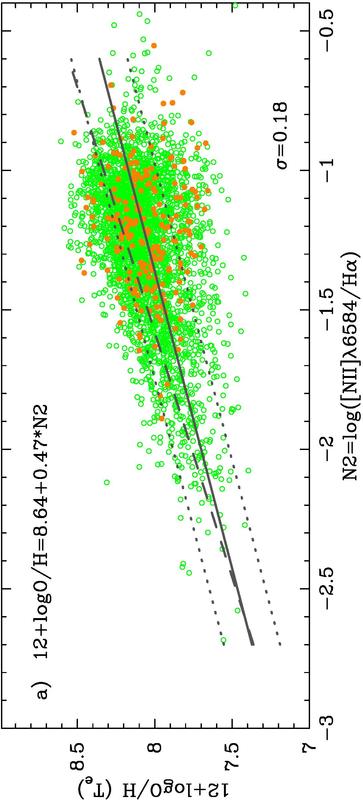}
\hspace{0.2cm}\includegraphics[angle=-90,width=0.48\linewidth]{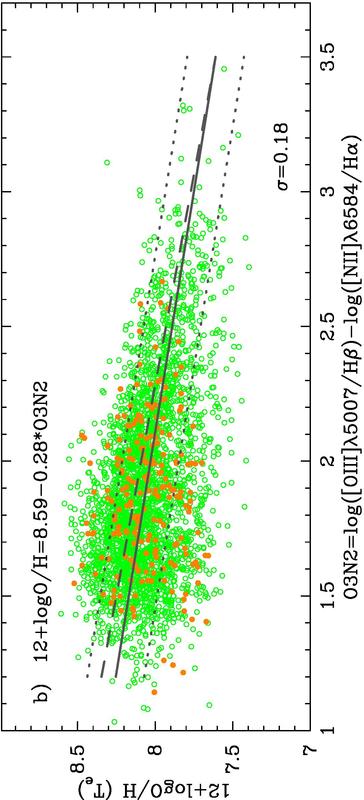}
}
\vspace{0.3cm}
\begin{center}
\hbox{
\includegraphics[angle=-90,width=0.48\linewidth]{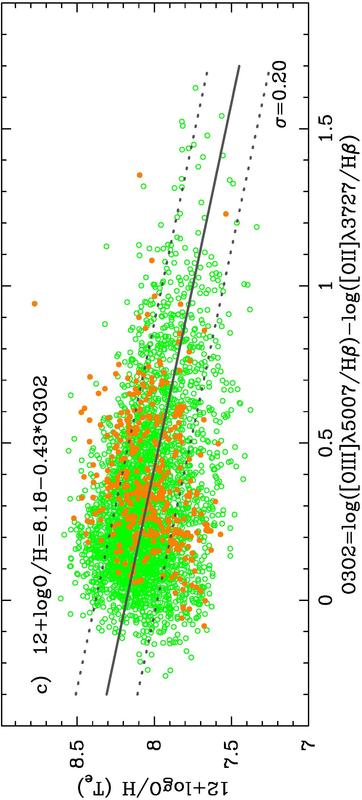}
\hspace{0.2cm}\includegraphics[angle=-90,width=0.48\linewidth]{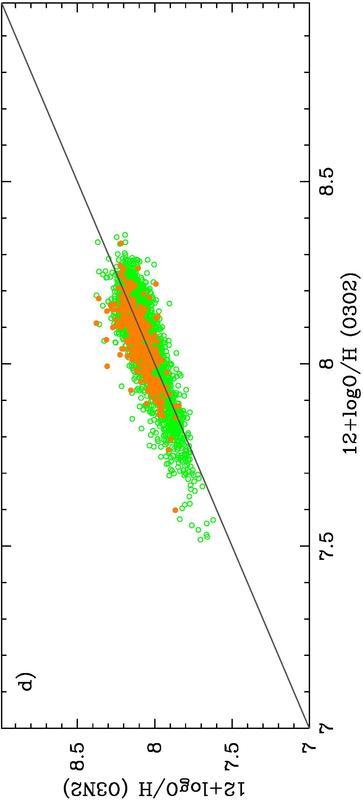}
}
\end{center}
\caption{ (a) - (c) Relations between the oxygen abundances 12+logO/H derived
by the $T_{\rm e}$-method and the line intensity ratios
[N~{\sc ii}]$\lambda$6584/H$\alpha$, 
[O~{\sc iii}]$\lambda$5007/[N~{\sc ii}]$\lambda$6584, and 
[O~{\sc iii}]$\lambda$5007/[O~{\sc ii}]$\lambda$3727, respectively. 
Shown are only the galaxies from the $T_{\rm e}$ sample.
Solid and dotted lines represent maximum-likelihood fits to the data and
1$\sigma$ dispersions, respectively, dashed lines in (a) and (b)
refer to the relations obtained by \citet{PP04}. (d) The relation between
the oxygen abundances derived using the $O3N2$ (Eq. \ref{eq:O3N2}) and $O3O2$ 
(Eq. \ref{eq:O3O2}) calibrations. The solid
line is the one-to-one correlation. Green open circles and brown filled circles
in all panels correspond to galaxies with the redshifts $z$ $\leq$ 0.3 and
$z$ $>$ 0.3, respectively.}
\label{fig6}
\end{figure*}

\section{Oxygen abundances}\label{sec:o}

One of the most reliable methods of the oxygen abundance determination
is a $T_{\rm e}$-method, which is based on the electron temperature derived
from the ([O~{\sc iii}]$\lambda$4959+$\lambda$5007)/$\lambda$4363
flux ratio. This method provides a good measure of the metallicity below 
about 12+logO/H $\la$ 8.3 \citep[e.g. ][]{Pi01,PP04,S05}. However, at 
super-solar metallicities the $T_{\rm e}$-method tends to underestimate 
significantly the true metallicity \citep{S05}. 

We apply the $T_{\rm e}$-method
for the $T_{\rm e}$ sample of the galaxies with
[O~{\sc iii}]$\lambda$4363 emission-line intensity measured at the level
better than 1$\sigma$. As it can be seen
below, all these galaxies satisfy condition 12+logO/H $\la$ 8.3.
   To determine oxygen abundances, we generally followed  
the procedures of \citet{ITL94,ITL97,I06}, which are
discussed in detail by e.g. \citet{I11a,I14a,I14b}.

Recently, \citet{P12} obtained new effective collision strengths
for the [O~{\sc iii}] transitions, which in the temperature
range 10000 - 20000K are higher by $\sim$20\% for the
transitions [O~{\sc iii}]$\lambda$4959, $\lambda$5007, but are
higher by only a few percent for the transition
[O~{\sc iii}]$\lambda$4363. These updates in the atomic data
would result in lower electron temperatures by $\sim$500 -- 700K
and hence in higher oxygen abundances \citep{N13} as compared to those obtained
with the \citet{I06} iterative procedure. However,
this effect will increase 12 + logO/H by not more than $\sim$ 0.05 dex.

\citet{N13} and \citet{Do13} considered the possibility that electrons
in H {\sc ii} regions may not be in thermal equilibrium, i.e. their
velocity distribution deviates from a Maxwellian distribution.
The overall effect will be an increase of the oxygen abundance. However,
for low-metallicity high-excitation H {\sc ii} regions selected in
this study the increase of 12 + logO/H is unlikely to be higher than 0.2 dex
\citep{Do13}.
We note that a deviation from the Maxwellian distribution
is found in the solar wind, but not in H {\sc ii} regions. 
Therefore, we use equations by \citet{I06}, which
assume a Maxwellian distribution of electrons.

The SDSS spectra of most star-forming galaxies are noisy. Then the 
[O~{\sc iii}]$\lambda$4363 line is not detected or is measured with 
insufficient 
accuracy. In these cases, strong-line methods are commonly used to derive oxygen
abundances instead of the $T_{\rm e}$-method. 
There are two different approaches to obtain calibrations for the metallicity
by using strong-line intensities. One approach is based on deriving
the line intensities by producing a grid of H~{\sc ii} region models
with varying 
metallicity, ionisation parameter, and other input parameters
\citep[e.g. ][]{CL01,KD02,KK04,D13}. The second approach uses the 
empirical
relations between certain combinations of strong-line intensities and
metallicity, which is derived by the $T_{\rm e}$-method 
\citep[e.g. ][]{A79,P79,Pi01,PP04,PT05,St06,P10}.

The advantage of all these SEL methods 
is that they are based on 
the strongest emission lines and thus can be applied to large galaxy samples.
The disadvantage is that most of these methods depend not only on the 
metallicity, but also on the ionisation parameter. 
The detailed comparison
of strong-line methods \citep[e.g. ][]{KD02,KE08,M08} shows that they 
are not consistent, non-monotonic \citep[e.g. the $R_{23}$-method,][]{P79,KD02}, 
depend on parameters 
different from metallicity and result in oxygen abundances, which can 
deviate by $\sim$ 0.5 dex or more.

To illustrate the problem we compare the oxygen abundances derived for
the $T_{\rm e}$ sample by the $T_{\rm e}$-method and different strong-line
calibrations by \citet{CL01}, which are based on population-synthesis
and H {\sc ii} region models. Calibrations by \citet{CL01} were used by
\citet{T04} to produce luminosity-metallicity and mass-metallicity
relations for a large sample of nearby SDSS star-forming galaxies. The 
\citet{T04} relations are often used as templates of the present-day 
luminosity-metallicity and mass-metallicity relations for comparison with
similar relations at higher redshifts.

We compare in Fig. \ref{fig4} oxygen abundances derived by using the 
$T_{\rm e}$-method and four out of seven strong-line calibrations 
by \citet{CL01}, 
which are based on different emission-line intensities and their ratios. 
The sample is split into two parts with the extinction-corrected ratio 
$O_{32}$ = 
[O~{\sc iii}]$\lambda$5007/[O~{\sc ii}]$\lambda$3727 $\geq$ 4.5 (small magenta 
filled circles) and $<$ 4.5 (small cyan open circles). The parameter 
$O_{32}$ is an increasing function 
of the ionisation parameter at given metallicity. Among the
\citet{CL01} calibrations the largest
number of emission lines ([O~{\sc ii}]$\lambda$3727, [O~{\sc iii}]$\lambda$5007,
[N~{\sc ii}]$\lambda$6584 and [S~{\sc ii}]$\lambda$6717, 6731) is used in
the calibration, which \citet{CL01} labelled as case~A. We do not consider 
two calibrations by \citet{CL01}, which use the blend of H$\alpha$ and
[N~{\sc ii}]$\lambda$6584 emission lines. 
This is because the spectral resolution
of the SDSS spectra is sufficient to separate these two lines.
We also excluded from the consideration one calibration, which is based on 
the discontinuity parameter $D_{4000}$ at $\lambda$ $\sim$ 4000\AA. This
calibration is suitable for galaxies with absorption lines in their spectra.

It is seen in Fig. \ref{fig4}a that the oxygen abundance at low metallicities 
(12 + logO/H ($T_{\rm e}$) $\la$ 8.1) derived by the strong-line method (case~A)
is by $\sim$ 0.2 -- 0.3 dex higher than that derived by the $T_{\rm e}$-method.
At higher metallicities the oxygen abundances derived by the two
methods are very different. The comparison of the $T_{\rm e}$-method
with other strong-line methods by \citet{CL01} 
(Fig.~\ref{fig4}b~--~\ref{fig4}d) also indicates large inconsistencies.

Therefore, we conclude that the oxygen abundances, and consequently,
the luminosity-metallicity and mass-metallicity relations
for nearby SDSS star-forming galaxies, obtained by \citet{T04}, may not be 
correct, because they use a set of \citet{CL01} internally inconsistent 
calibrations.

A more promising way is to use empirical relations between line intensities 
and oxygen abundances derived by the $T_{\rm e}$-method in real star-forming
galaxies \citep[e.g. ][]{PP04}. These relations can be obtained only for 
galaxies with 
high-excitation H~{\sc ii} regions, where the [O~{\sc iii}]$\lambda$4363
emission line can be detected. Our $T_{\rm e}$ sample satisfies these conditions.
In Fig. \ref{fig5} we compare oxygen abundances obtained by the 
$T_{\rm e}$-method with those, which were obtained by using strong-line 
calibrations by \citet{PP04}. It is seen that 
abundances derived by different methods are in agreement despite the
larger dispersion and small offset of $\sim$ 0.1 at 12+logO/H $\geq$ 8.0. 

Using only SDSS data for the large $T_{\rm e}$ sample
we modified the \citet{PP04} calibrations (Fig. \ref{fig6}) and obtained
the maximum-likelihood linear relations
\begin{equation}
{\rm 12+log\frac{O}{H}}= 8.64\pm0.02 + (0.47\pm0.02)\times N2, \label{eq:N2}
\end{equation}
\begin{equation}
{\rm 12+log\frac{O}{H}}= 8.59\pm0.02 - (0.28\pm0.01)\times O3N2, \label{eq:O3N2}
\end{equation}
where $N2$ = log([N~{\sc ii}]$\lambda$6584 / H$\alpha$),
$O3N2$ = log([O~{\sc iii}]$\lambda$5007 / H$\beta$) -- 
log([N~{\sc ii}]$\lambda$6584 / H$\alpha$).

It is seen in Fig. \ref{fig6} that relations by \citet{PP04} 
(dashed lines) obtained for a smaller sample of H~{\sc ii} regions, 
but for a larger range of oxygen abundances with inclusion of
high-metallicity H~{\sc ii} regions, are not very different from
the relations Eqs. \ref{eq:N2} and \ref{eq:O3N2} (solid lines) obtained in this
paper. However, the difference is smaller ($\leq$0.1 for 12 + logO/H)
in the case of $O3N2$ calibrations (Fig. \ref{fig6}b).
The galaxies from our entire sample
were selected using the same selection criteria, as the 
$T_{\rm e}$ sample, i.e. H~{\sc ii} regions in these galaxies
are of high excitation. The only difference is that the 
[O~{\sc iii}]$\lambda$4363 was measured with insufficient
accuracy or not detected in many galaxies. Therefore,
calibrations Eqs. \ref{eq:N2} and \ref{eq:O3N2} can be applied
for the entire sample.

\begin{figure}
\hbox{
\includegraphics[angle=-90,width=0.98\linewidth]{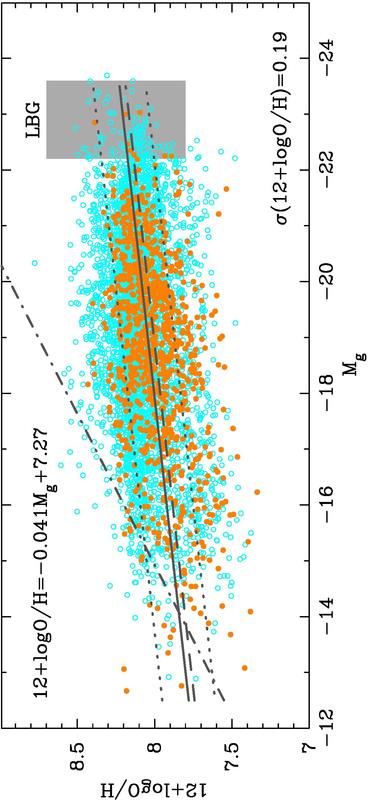}
}
\caption{The relations between extinction-corrected absolute SDSS $g$-band 
magnitude $M_g$ and oxygen abundance 12+logO/H. 
The galaxies with the [O~{\sc iii}]$\lambda$4363 emission-line detection 
at a level higher than 4$\sigma$ shown by brown 
filled circles and 
the rest of the sample by cyan open circles.
The solid line provides a linear maximum-likelihood fit to the data and 
dotted lines show 1$\sigma$ dispersions. The dashed line is a linear 
maximum-likelihood fit to the data shown by brown symbols.
The steep dash-dotted 
line indicates the relation by \citet{T04}.
The location of Lyman-break galaxies (LBGs) at $z$ $\sim$ 3 
by \citet{P01} is marked by a grey rectangle.
}
\label{fig7}
\end{figure}

However, there are 638 galaxies from the entire sample,
including 264 galaxies from the $T_{\rm e}$ sample, where the
[N~{\sc ii}]$\lambda$6584
emission line has not been measured because of its weakness or 
because this line is outside the SDSS wavelength range in high-redshift galaxies
with $z$ $\ga$ 0.40 of the SDSS DR7 survey and $z$ $\ga$ 0.55 of
the SDSS/BOSS survey. In all these cases, with exception of the galaxies
from the $T_{\rm e}$ sample, we use the relation
between the oxygen abundance 12+logO/H and 
the parameter $O3O2$ = log([O~{\sc iii}]$\lambda$5007/[O~{\sc ii}]$\lambda$3727)
derived for the $T_{\rm e}$ sample (Fig. \ref{fig6}c)
\begin{equation}
{\rm 12+log\frac{O}{H}}= 8.18\pm0.01 - (0.43\pm0.02)\times O3O2. \label{eq:O3O2}
\end{equation}

We compare in Fig. \ref{fig6}d oxygen abundances derived from
Eqs. \ref{eq:O3N2} and \ref{eq:O3O2}. It is seen that 12 +logO/H obtained
with the both $O3N2$ and $O3O2$ calibrations are nearly identical.
Furthermore, in all panels of Fig. \ref{fig6} there is no evident offset 
between the galaxies with low redshifts $z$ $\leq$ 0.3 (green open circles)
and higher redshifts $z$ $>$ 0.3 (brown filled circles), implying that all
three relations Eqs. \ref{eq:N2} -- \ref{eq:O3O2} can be applied in the entire
range of redshifts.

Summarising, we derive oxygen abundances (1) by the $T_{\rm e}$-method
in 3607 galaxies, (2) by using Eq. \ref{eq:O3O2} in 638 -- 264 = 374 
galaxies, and (3)
by using Eq. \ref{eq:O3N2} in the remaining galaxies.

\section{Results \label{res}}

\subsection{Absolute $g$-band magnitude - metallicity relation}

The extinction-corrected absolute $g$-band magnitude - metallicity relation for 
the entire sample of star-forming galaxies is shown in Fig. \ref{fig7}. 
The maximum-likelihood linear regression,
\begin{equation}
12+\log \frac{\rm O}{\rm H} = -(0.041\pm0.002) M_g + 7.27\pm0.03, \label{Mg_o}
\end{equation}
is shown by a solid line and 1$\sigma$ deviations by dotted lines.
For comparison, we show by dashed line the maximum-likelihood linear 
regression for
the $T_{\rm e}$ subsample with the [O~{\sc iii}]$\lambda$4363 emission line 
detected at the level higher than 4$\sigma$. The difference between these two 
relations is very small, less than 0.05 in 12 + logO/H. 
This indicates that decreasing the detection limit
for the [O~{\sc iii}]$\lambda$4363 emission line to 1$\sigma$ and including
galaxies with 12 + logO/H derived by $O3N2$ or $O3O2$ calibrations
leads to a higher dispersion by a factor of $\sim$2 but does not introduce 
significant offsets in 12 + logO/H.

The relation in Eq. \ref{Mg_o} is slightly flatter than that obtained 
by \citet{G09} and \citet{I11a}. 
However, we note that \citet{G09} and \citet{I11a}
considered not only SDSS galaxies but also galaxies from other samples
with a large fraction of low-metallicity galaxies.
Our sample is more uniform and includes 
only SDSS compact galaxies. The $M_g$ -- 12+logO/H relation for this sample
is very similar to that derived by \citet{I14a}
for a much smaller sample of SDSS compact star-forming galaxies.

\begin{figure*}
\hbox{
\includegraphics[angle=-90,width=0.48\linewidth]{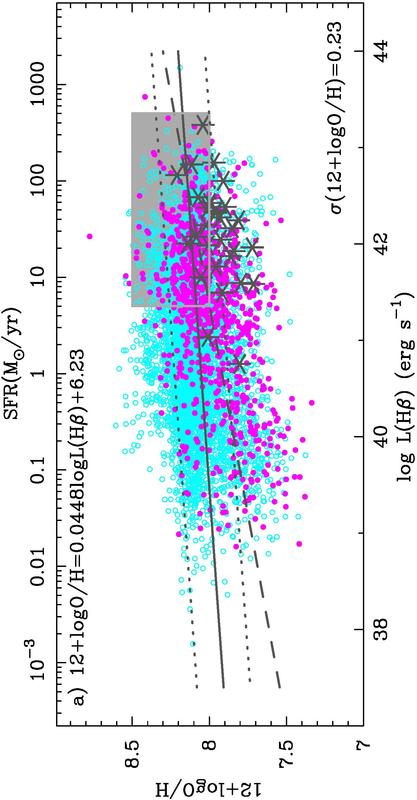}
\includegraphics[angle=-90,width=0.48\linewidth]{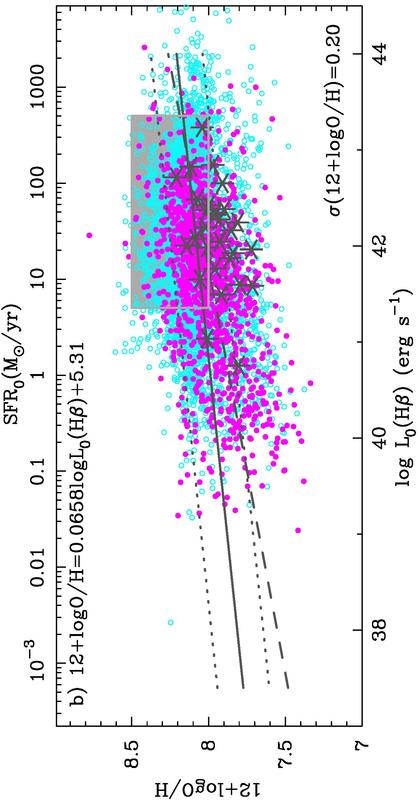}
}
\caption{The H$\beta$ luminosity-metallicity relation. The upper axis
indicates the scale of the star-formation rate, calculated with \citet{K98}
calibration. Galaxies with the equivalent widths EW(H$\beta$) $<$100\AA\ 
and $\geq$100\AA\ are shown by the open cyan circles and filled magenta 
circles, respectively. 
The H$\beta$ luminosities in panel (a) are corrected
for extinction and aperture, while they are additionally reduced to zero 
starburst age in panel (b). Solid lines in both panels are maximum-likelihood 
linear fits to all data and dotted lines are 1$\sigma$ dispersions.
The dashed lines are maximum-likelihood linear fits only for the 
galaxies with EW(H$\beta$) $\geq$100\AA\ shown by magenta symbols.
The location of star-forming galaxies with redshifts 2.0 $<$ $z$ $<$ 2.6
by \citet{S14} and with redshifts $z$ $>$ 2 by \citet{C14} is shown by grey 
rectangles. Star-forming galaxies with redshifts $z$ $\sim$ 3.4 by 
\citet{T14} are indicated by dark asterisks.
}
\label{fig8}
\end{figure*}

Our $g$-band absolute magnitude - metallicity
relation is much flatter than 
that obtained by \citet{T04} for a sample of $\sim$ 53000 star-forming 
galaxies selected from the SDSS (dash-dotted line in Fig. \ref{fig7}). 
This difference arises primarily because of different methods applied for
the metallicity determination and because of the selection criteria used 
in both studies.
Our sample consists of compact galaxies with high-excitation H~{\sc ii}
regions, and
we applied strong-line methods, which give oxygen abundances consistent 
with the $T_{\rm e}$-method, at variance with \citet{CL01} calibrations used by
\citet{T04}. Furthermore, most of the star-forming galaxies from the \citet{T04}
sample are galaxies with low-excitation H~{\sc ii} regions, for which
the determination of metallicities is more uncertain compared to
galaxies with high-excitation H~{\sc ii} regions. Since 
empirical calibrations inferred from observations are not available
for low-excitation H~{\sc ii} regions because of the very weak and undetectable
[O~{\sc iii}]$\lambda$4363 emission lines, we do not consider these galaxies.

On the other hand, the location
of star-forming galaxies at 0~$<$~$z$~$<$~1 with high luminosities
is similar to that for
the Ly-break galaxies (LBGs) by \citet{P01}, shown by a box in Fig.~\ref{fig7}.
This is consistent with a universal character of the $M_g$~--~12+logO/H 
relation for star-forming galaxies with high-excitation H~{\sc ii} regions 
in the entire redshift range  0~$<$~$z$~$<$~3.

\subsection{H$\beta$ luminosity - metallicity relation}

\begin{figure*}
\hbox{
\includegraphics[angle=-90,width=0.48\linewidth]{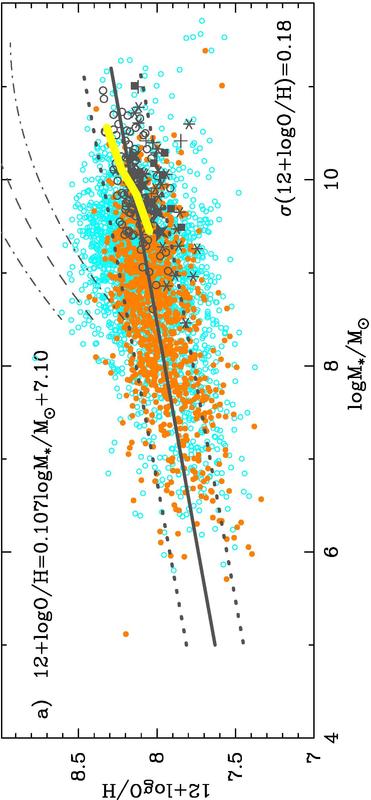}
\hspace{0.2cm}\includegraphics[angle=-90,width=0.48\linewidth]{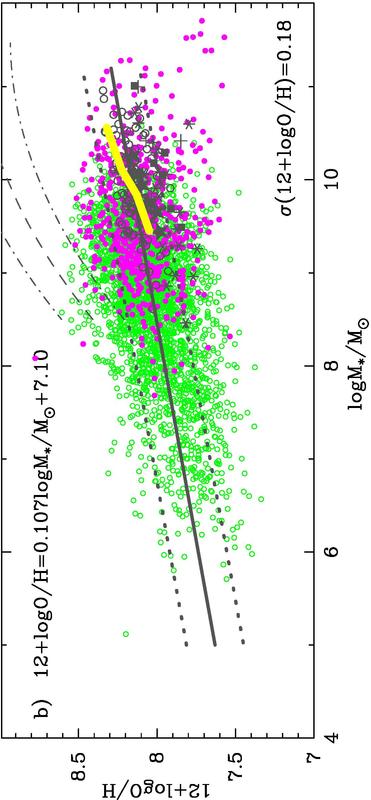}
} 
\vspace{0.3cm}
\hbox{
\includegraphics[angle=-90,width=0.48\linewidth]{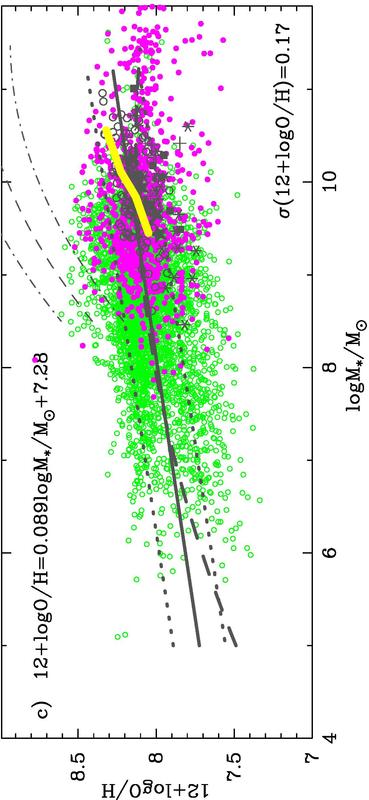}
\hspace{0.2cm}\includegraphics[angle=-90,width=0.48\linewidth]{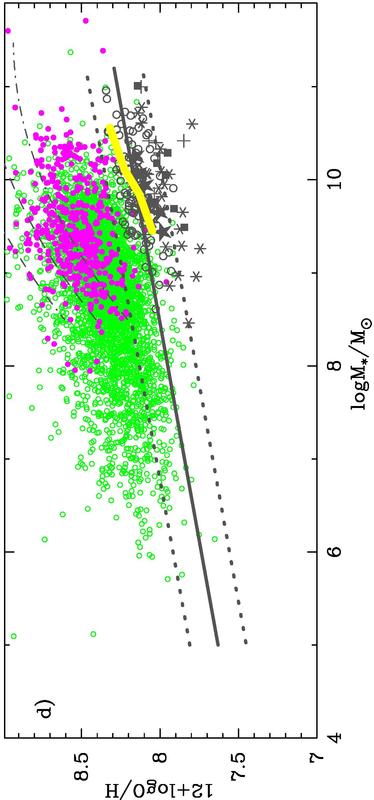}
} 
\caption{ The stellar mass - metallicity relations for (a) -- (b) 
the $T_{\rm e}$ and (c) -- (d) the entire sample. 
Brown symbols in (a) correspond to the galaxies, where 
[O~{\sc iii}]$\lambda$4363 emission is detected at a level better than 
4$\sigma$, while cyan symbols are for the rest of the galaxies. 
Galaxies shown with magenta and green symbols in (b) -- (d) have 
SFR $\geq$ 10 $M_\odot$ yr$^{-1}$ and $<$ 10 $M_\odot$ yr$^{-1}$,
respectively. Thick solid lines in (a) -- (c) represent maximum-likelihood 
linear fits 
to the data in respective panels and dotted lines indicate 1$\sigma$ 
dispersions. The thick dashed line in (c) is the maximum-likelihood 
quadratic fit to the data.
The thick solid line in (d) is the maximum-likelihood linear fit 
to the data in (a).
The relation by \citet{T04} is shown by thin dashed lines
and dot-dashed lines are 1$\sigma$ dispersions. 
The location of stacked
data for star-forming galaxies at $z$ $>$ 2 by \citet{C14} and of
data for ninety six star-forming galaxies at 2.0 $<$ $z$ $<$ 2.6 by \citet{S14} are
shown by dark-grey filled stars and open circles, respectively. 
Star-forming galaxies at $z$ $\sim$ 3.5 by \citet{M08} and at $z$ $\sim$ 2
by \citet{M14} are shown by dark-grey filled squares and crosses, respectively. 
Thirty star-forming galaxies with redshifts 
$z$ $\sim$ 3.4 by \citet{T14} are indicated by asterisks. 
The relation for $z$ $\sim$ 2.3 by \citet{San15} is shown by thick yellow solid
line. Oxygen abundances
for all high-$z$ galaxies are calculated with $O3N2$ or $O3O2$ 
calibrations from this paper (Eqs. \ref{eq:O3N2} -- \ref{eq:O3O2}).
(d) Same as in (c), but oxygen abundances are derived
using the \citet{CL01} strong-line method (case A).
}
\label{fig9}
\end{figure*}

H$\beta$ emission is a signature of the youngest stellar population in 
star-forming galaxies and the H$\beta$ luminosity allows us to estimate the
SFR using, e.g., the relation by \citet{K98} between SFR and H$\alpha$ 
luminosity. However, we note
that the use of the relation by \citet{K98} is not well justified for
starburst galaxies because of uncertainties in the adopted duration of the
burst. The H$\beta$ and H$\alpha$ luminosities are more reliable
characteristics, because they are directly derived from observations.

The UV, H$\beta$ and mid-infrared luminosities of compact
star-forming galaxies are tightly correlated \citep[e.g. ][]{I14a} indicating
that the radiation of young stellar populations is a dominant source of
dust heating and of galaxy emission in the UV range and H$\beta$ line.
Consequently, SFRs derived by different methods for star-forming
galaxies in a wide range of redshifts can be compared.

In Fig. \ref{fig8}a we show the extinction- and aperture-corrected H$\beta$
luminosity -- oxygen abundance relation for our entire sample. The upper axis
denotes the SFR calculated with the \citet{K98} relation. The box indicates the
location of star-forming galaxies with redshifts 2.0~$<$~$z$~$<$~2.6 by
\citet{S14} and with $z$ $>$ 2 by \citet{C14}. Star-forming galaxies
with $z$ $\sim$ 3.4 by \citet{T14} are indicated by asterisks. Strong 
emission lines with [O~{\sc iii}]$\lambda$4959/H$\beta$ $\ga$ 1 are present in 
spectra of most galaxies from their samples. Therefore, strong-line
methods calibrated with the $T_{\rm e}$-method can be applied. 
\citet{S14} compared SFRs obtained
for their sample from the UV continuum luminosities and H$\alpha$ luminosities
and found good agreement. They derived oxygen abundances by using 
strong-line $O3N2$ calibration, similar to that applied for our sample. 
To derive
metallicities \citet{C14} and \citet{T14} used empirical calibrations 
by \citet{M08}, which also were calibrated with the use of galaxy samples, where
oxygen abundances were derived by the $T_{\rm e}$-method. However, 
for consistency reasons we selected only high-$z$ galaxies
with [O~{\sc iii}] $\lambda$4959/H$\beta$ $\geq$ 1 and recalculated their
oxygen abundances using Eqs. \ref{eq:O3N2} or \ref{eq:O3O2}.

Despite the noisy data for high-$z$ galaxies there is a good coincidence 
in Fig. \ref{fig8}a between the luminous tail in the distributions of our 
galaxies and the high-$z$ galaxies, 
indicating the universal character of the $L$(H$\beta$) -- 12+logO/H 
relation for star-forming galaxies with high-excitation H~{\sc ii} regions.

However, we note the offset between the distributions of the galaxies with
EW(H$\beta$) $\geq$ 100\AA\ (magenta symbols and dashed line) and of 
the galaxies with EW(H$\beta$) $<$ 100\AA\
(cyan symbols and solid line).
This difference is primarily due to the fact that starbursts with high 
EW(H$\beta$) are on average younger. The H$\beta$ emission is powered by the
ionising radiation and its luminosity strongly declines with the age of
starburst. According to the Starburst99 models \citep{L99} the number of 
ionising photons and, respectively, the H$\beta$ luminosity 
produced by a burst is constant during the first 3 Myr. After that time it
declines and is lower by a factor of $\sim$ 50 at a starburst age
of 10 Myr. On the other hand, the luminosity of non-ionising radiation 
at $\sim$1500\AA\ during the same time interval is decreased only by a factor
of $\sim$ 2. Since galaxies are observed with different
starburst ages, their H$\beta$ luminosities, for consistency, should be reduced
to zero age. The H$\beta$ equivalent width can be used for this, because
it depends on the starburst age. We
use Starburst99 models to derive the correction of the H$\beta$ luminosity
for the starburst age,
\begin{equation}
\Delta \log L({\rm H}\beta) = 2.553 - \log [{\rm EW}({\rm H}\beta)]  \label{eq:Hbcorr}
\end{equation}
for $\log {\rm EW}({\rm H}\beta) \leq 2.553$, otherwise 
$\Delta \log L({\rm H}\beta) = 0$. The relation Eq. \ref{eq:Hbcorr}
is correct for instantaneous bursts and may overestimate the correction 
of the H$\beta$ luminosity for the starburst age
if underlying older stellar population is present. Our SED fitting shows 
that the contribution
of an older stellar population to the stellar 
continuum near the H$\beta$ emission
line in galaxies with strong emission lines varies in a wide range of 
$\sim$ 5\% -- 50\%. This results in an overestimation of 
$\log L({\rm H}\beta)$ by $\la$ 0.3. The effect is stronger for galaxies
with low EW(H$\beta$).

In Fig. \ref{fig8}b we show the relation between the H$\beta$ luminosity
$L_0$(H$\beta$) reduced to zero age and the oxygen abundance.
It is seen that the differences between maximum-likelihood linear fits for our 
two samples (solid and dashed lines) are reduced and distributions of data
are in better agreement compared to  Fig. \ref{fig8}a, 
implying that the age correction should be applied for 
consistent comparison of H$\beta$ and H$\alpha$ luminosities and
star-formation rates in star-forming galaxies.

 \begin{table*}
 \caption{Average oxygen abundances for the entire sample in the 
four stellar mass intervals}
 \label{tab1}
 \begin{tabular}{ccccc} \hline
SFR        &\multicolumn{4}{c}{12+logO/H$^{\rm a}$} \\ \cline{2-5}
($M_\odot$yr$^{-1}$)&9.0 -- 9.2$^{\rm b}$   &9.4 -- 9.6$^{\rm b}$  &10.0 -- 10.2$^{\rm b}$  &10.4 -- 10.6$^{\rm b}$ \\ \hline
  0.1 - 1         &8.10$\pm$0.18 (205)  &8.12$\pm$0.16 (142)  & ...                 & ...                  \\
  1 - 5           &8.10$\pm$0.16 (147)  &8.14$\pm$0.15 (163)  &8.12$\pm$0.16 (~\,25)&8.14$\pm$0.09 (~\,~\,6) \\
  5 - 10          &8.11$\pm$0.15 (~\,67)&8.14$\pm$0.13 (~\,69)&8.12$\pm$0.15 (~\,24)&8.15$\pm$0.07 (~\,~\,4) \\
 10 - 20          &8.12$\pm$0.16 (~\,52)&8.12$\pm$0.20 (~\,42)&8.15$\pm$0.16 (~\,30)&8.09$\pm$0.13 (~\,~\,5) \\
 20 - 50          &8.12$\pm$0.15 (~\,27)&8.07$\pm$0.18 (~\,46)&8.13$\pm$0.11 (~\,28)&8.09$\pm$0.21 (~\,15) \\
  $>$ 50          &     ...             &8.08$\pm$0.15 (~\,22)&7.98$\pm$0.20 (~\,14)&8.12$\pm$0.13 (~\,~\,9) \\
 \hline
 \end{tabular}

$^{\rm a}$The number of galaxies used for averaging is shown in parentheses.\\
$^{\rm b}$ interval of log $M_*/M_\odot$.
 \end{table*}

\begin{figure*}
\hbox{
\hspace{1.5cm}\includegraphics[angle=-90,width=0.80\linewidth]{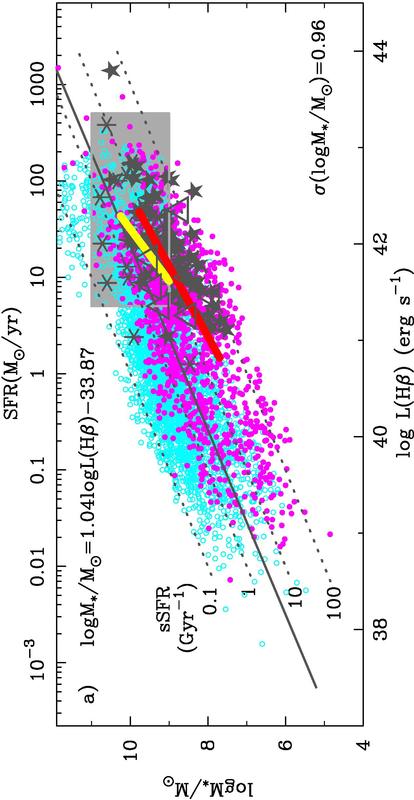}}
\vspace{0.3cm}
\hbox{
\hspace{1.5cm}\includegraphics[angle=-90,width=0.80\linewidth]{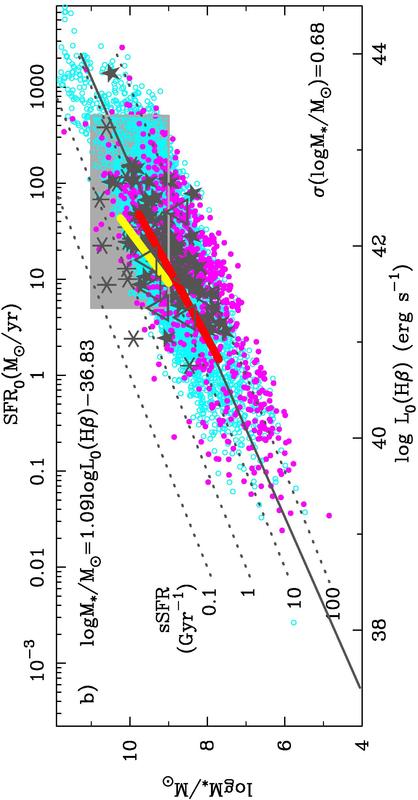}}
\caption{The stellar mass - H$\beta$ luminosity (or equivalently 
star-formation rate) relations for the entire sample. The extinction-
and aperture-corrected H$\beta$ luminosities and stellar masses 
are shown in panel (a), while 
H$\beta$ luminosities are additionally reduced to zero starburst age in 
panel (b).  Galaxies with the equivalent widths EW(H$\beta$) $<$100\AA\ 
and $\geq$100\AA\ are shown by the small cyan open circles and small magenta 
filled circles, respectively. Solid lines
represent linear maximum-likelihood fits to all data. Constant specific 
star-formation
rates (sSFR=SFR/$M_*$) are shown by dotted lines and labelled by values in units
of Gyr$^{-1}$.
The location of star-forming galaxies with redshifts 2.0 $<$ $z$ $<$ 2.6
by \citet{S14} and with $z$ $>$ 2 by \citet{C14} is shown by a grey rectangle. 
Ly$\alpha$ emitting galaxies 
with redshifts 1.9 $<$ $z$ $<$ 3.6 by \citet{H14} and star-forming
galaxies with redshifts 
$z$ $\sim$ 3.4 by \citet{T14} are shown by dark-grey stars and asterisks, 
respectively. The relations for galaxies with $z$ $\sim$ 4 by \citet{B12} and
for galaxies with $z$ $\sim$ 5 -- 6 by \citet{Sal15} are indicated by thick
solid red and yellow lines, respectively. The galaxy candidates with
$z$ $\sim$ 9 -- 11 by \citet{C13} and \citet{O14} are shown by large open 
dark-grey triangles. The H$\beta$ luminosities
of all high-$z$ galaxies in (b) are not reduced to zero starburst age.
}
\label{fig10}
\end{figure*}

\subsection{Stellar mass - metallicity relation}

In Fig. \ref{fig9}a we show the extinction- and 
aperture-corrected stellar mass - metallicity relations
for the $T_{\rm e}$ sample. The sample is split into two subsamples
with [O~{\sc iii}]$\lambda$4363 emission line detected at the level better than
4$\sigma$ (small filled brown circles) and in the range 
1$\sigma$ -- 4$\sigma$ (small open cyan circles).
The data for the $T_{\rm e}$ sample can be fitted 
by a linear relation, as shown in Fig. \ref{fig9}a
by the solid line,
\begin{equation}
12 + \log \frac{\rm O}{\rm H} = 
(0.107\pm0.004) \log \frac{M_*}{M_\odot}  + 7.10\pm0.04. \label{mtot_o1}
\end{equation}

The dispersion for the subsample with weaker [O~{\sc iii}]$\lambda$4363 lines
is characterised by a factor of $\sim$2 higher value, but no offset relative to
the subsample with stronger lines is present. Similarly, the subsample
with high star-formation rate SFR $\geq$ 10 $M_\odot$ yr$^{-1}$ 
(small filled magenta circles in Fig. \ref{fig9}b) follows the same relation
as the subsample with low  SFR $<$ 10 $M_\odot$ yr$^{-1}$ (small open green 
circles), indicating that there is no obvious 
dependence of the metallicity on SFR.

In Fig. \ref{fig9}c the stellar mass -- metallicity relation is shown for
the entire sample. Adding the galaxies with the oxygen abundance derived with
strong-line methods results in a flattening of the relation at 
$M_*$ $\geq$ 10$^{10}$ $M_\odot$. Linear regression yields
\begin{equation}
12 + \log \frac{\rm O}{\rm H} = 
(0.089\pm0.004) \log \frac{M_*}{M_\odot}  + 7.28\pm0.04. \label{mtot_o2}
\end{equation}
However, better agreement can be obtained with the quadratic 
stellar mass -- metallicity relation 
\begin{equation}
12 + \log \frac{\rm O}{\rm H} = 
\log \frac{M_*}{M_\odot}\left(0.023 \log \frac{M_*}{M_\odot} + 0.466\right)  + 5.72 \label{mtot_o2_1}
\end{equation}
shown in Fig. \ref{fig9}c by a thick dashed line.
These relations are similar to the
relation obtained by \citet{I14a} for a smaller sample of SDSS galaxies.

On the other hand, the stellar mass-metallicity relations in Fig.~\ref{fig9}a --
\ref{fig9}c (solid lines)
are much flatter and are shifted to lower metallicities 
compared to that obtained by \citet{T04} (thin dashed lines).
In part, this difference may be caused by different selection criteria.
However, we argue that the main cause of the difference is the use of different 
methods for the oxygen abundance determination, as discussed above.
We show in Fig. \ref{fig9}d the stellar mass -- metallicity relation for
the entire sample, same as in Fig. \ref{fig9}c, but with oxygen abundances 
derived by the \citet{CL01} calibration, which has been used by 
\citet{T04}. It is seen that agreement with the \citet{T04}
relation is much better if the \citet{CL01} calibration is used. However,
this calibration is not consistent with the $T_{\rm e}$-method, as it was 
discussed above.

It was argued in some recent papers \citep[e.g. ][]{M10,L10,H12,AM13} 
that star-forming galaxies with higher SFRs are systematically more
metal-poor on the stellar mass-metallicity diagrams. 
In particular, \citet{M10} proposed a fundamental metallicity relation (FMR)
between stellar mass, gas-phase metallicity and star formation rate.

We checked whether the tendency of lower metallicity in galaxies
with higher SFR is present for galaxies from our sample. To exclude the 
dependence of the oxygen abundance on the stellar
mass we calculated average 12 + logO/H for galaxies from the entire sample 
in the four narrow ranges of stellar masses and
for different ranges of SFR (Table \ref{tab1}). 
We find no evidence that the galaxies with high SFRs 
are systematically more metal-poor, 
confirming conclusions by \citet{I14a}.
We also applied the $\chi^2$ minimization technique for samples shown in
Fig. \ref{fig9}b in the plane [$\mu_\alpha$ = log($M_*/M_\odot$) -- 
$\alpha$logSFR($M_\odot$yr$^{-1}$)] -- metallicity and find $\alpha$ $\sim$ 0, 
i.e. no dependence on SFR is detected.

\citet{M10} argued that introducing $\mu_\alpha$ with $\alpha$ = 0.32
eliminates any redshift evolution up to $z$ $\sim$ 2.5 yielding
the same $\mu_\alpha$ -- metallicity relation independent of redshift 
(see also Sect. \ref{intro}).
To verify this conclusion with our sample we show in Fig. \ref{fig9}
star-forming galaxies at $z$ $>$ 2 by \citet{C14} (stars), star-forming
galaxies at 2.0 $<$ $z$ $<$ 2.6 by \citet{S14} (filled circles), star-forming 
galaxies at redshifts $z$ $\sim$ 3.5 by \citet{M08} (filled squares), 
star-forming galaxies with $z$ $\sim$ 2 by \citet{M14} (crosses), 
and star-forming galaxies with $z$ $\sim$ 3.4 by \citet{T14} (asterisks). The
relation for galaxies with $z$ $\sim$ 2.3 by \citet{San15} is shown
by thick yellow solid line. 
For homogeneity, only high-$z$ galaxies with 
[O~{\sc iii}]$\lambda$4959/H$\beta$ $\geq$ 1 and oxygen abundances derived by
Eqs. \ref{eq:O3N2} or \ref{eq:O3O2} are shown.
It is seen that high-$z$ galaxies 
are closely related to 
lower-$z$ counterparts and do not indicate any appreciable dependence of the 
mass -- metallicity relation on redshift. 
However, we note that the high-$z$ data by \citet{M08}, \citet{T14}, and 
\citet{M14} are offset to lower metallicities compared to the data 
by \citet{C14} and \citet{S14}. This offset may indicate
some evolution with the redshift, because \citet{M08} and \citet{T14}
considered galaxies at higher redshifts \citep[but not ][]{M14}.
The mass-metallicity relation for high-$z$ galaxies is considerably
flatter and is in better agreement with the relation for low-$z$ galaxies
in the mass range log~$M_*/M_\odot$ = 8.5 - 11,
if only data by \citet{C14} and \citet{S14} are considered.

\subsection{H$\beta$ luminosity - stellar mass relation}

Finally, we consider the relation between the H$\beta$ luminosity 
(or, equivalently, the SFR) and stellar mass, both corrected for extinction and
aperture. At variance with relations discussed above this relation does not
depend on the technique used for the determination of metallicity.
Fig. \ref{fig10}a shows the relation between the extinction- and 
aperture-corrected H$\beta$ luminosities (or SFRs) and stellar masses 
for the entire sample. The linear maximum-likelihood fit to the data is
shown by a solid line. The sample is characterised by very high specific
star formation rates sSFR = SFR/$M_*$ ranging from 
$\sim$ 0.1 Gyr$^{-1}$ to $\sim$ 100 Gyr$^{-1}$ (dotted lines), 
which are among the highest for the local and high-$z$ star-forming 
galaxies. It is clearly seen that galaxies
with younger starbursts (magenta symbols) have systematically higher 
H$\beta$ luminosities, as compared to the galaxies with older starbursts 
(cyan symbols).

For comparison, in Fig. \ref{fig10}a we show data for high-$z$ star-forming
galaxies by \citet{C14} and \citet{S14} (inside the box), Ly$\alpha$-emitting 
galaxies by \citet{H14} (stars), star-forming galaxies by \citet{T14} 
(asterisks), star-forming galaxies with $z$ $\sim$ 4 by \citet{B12} (thick
red solid line), star-forming galaxies with $z$ $\sim$ 5 -- 6
by \citet{Sal15} (thick yellow solid line), and galaxy candidates with
$z$ $\sim$ 9 -- 11 by \citet{C13} and \citet{O14} (large open dark-grey 
triangles). 
High-$z$ galaxies, including the highest redshift galaxy candidates, 
nicely follow the trend for low-$z$ galaxies with 
EW(H$\beta$) $\geq$100\AA\ (magenta symbols), 
but are offset to higher $L$(H$\beta$) as compared to the rest of galaxies
from our sample with presumably older starbursts (cyan symbols).

Reducing the H$\beta$ luminosities (or, equivalently, SFRs) of galaxies
from our sample to zero starburst age results in a much tighter relation in the
redshift range 0 $<$ $z$ $<$ 1 (magenta and cyan symbols in Fig. \ref{fig10}b). This is because the age correction
for galaxies with EW(H$\beta$) $\geq$ 100\AA\ is much smaller compared to that
for the remaining galaxies of our sample.
We note that the correction for the starburst age of the
stellar mass is small because this parameter was obtained from the SED in
the optical range, which is insensitive to age variations on the scale
of several Myr. Therefore we did not introduce a correction for the stellar 
mass.

It is seen in Fig. \ref{fig10}b that the distributions of low-$z$ and high-$z$
star-forming galaxies in the $L_0$(H$\beta$) -- $M_*$/$M_\odot$ plane are
very similar, implying that the physical properties of these galaxies are 
similar as well. However, we note that H$\beta$ luminosities for 
high-$z$ galaxies are not corrected for the starburst age.

\section{Summary \label{sum}}

We studied the relations between global parameters (absolute magnitudes, 
H$\beta$ luminosities, star-formation rates SFR, stellar masses, and
gas-phase oxygen abundances) of 5182 compact star-forming
galaxies with high-excitation H~{\sc ii} regions 
in the redshift range 0 $<$ $z$ $<$ 1 selected from the SDSS DR7 and 
SDSS/BOSS DR10 surveys. These data were split into two samples, one is
the $T_{\rm e}$ sample of 3607 galaxies, where the temperature-sensitive 
[O~{\sc iii}]$\lambda$4363 emission line was detected at the level better 
than 1$\sigma$ and the oxygen abundance was derived by the $T_{\rm e}$-method,
and the second sample consisting of the remaining galaxies, where the oxygen 
abundances were derived by strong-line methods.
The data for local galaxies were combined with
respective literature data for high-$z$ galaxies (with 2 $\la$ $z$ $\la$ 3)
to detect a possible redshift evolution of global parameters in
star-forming galaxy.

Our main results are as follows.

1. Using the $T_{\rm e}$ sample we analysed the oxygen abundances derived
by the $T_{\rm e}$-method and various strong-emission-line (SEL) 
methods used for the chemical abundance determination. We adopted 
strong-line methods, which are consistent with the $T_{\rm e}$-method
for high-excitation H~{\sc ii} regions.

2. We analysed various relations between absolute magnitudes, 
H$\beta$ luminosities (or equivalently star-formation rates SFR), 
stellar masses, and oxygen abundances for our sample of low-$z$
compact star-forming galaxies in combination with the respective
data for high-$z$ star-forming galaxies. We find good agreement
between relations for low-$z$ and high-$z$ galaxies indicating very
weak redshift evolution of global parameters and weak dependence
of metallicity on SFR, contrary to results in some other studies.
This finding emphasizes the universal character of relations for 
considered global parameters of compact star-forming galaxies 
with high-excitation H {\sc ii} regions at 
redshifts 0~$\la$~$z$~$\la$~3. Furthermore, galaxy candidates at 
$z$ $\sim$ 9 - 11 do not deviate from low-$z$ compact star-forming galaxies
on the $M_*$ -- SFR diagram implying that universal relations
between global galaxy parameters may be valid for galaxies at higher redshifts.

\section*{Acknowledgements}
Y.I.I., N.G.G. and K.J.F. are grateful to the staff of the Max Planck 
Institute for Radioastronomy (MPIfR) for their warm hospitality. 
Y.I.I. and N.G.G. acknowledge financial support by the MPIfR.
Funding for the SDSS and SDSS-II has been provided by the Alfred P. Sloan Foundation, the Participating Institutions, the National Science Foundation, the U.S. Department of Energy, the National Aeronautics and Space Administration, the Japanese Monbukagakusho, the Max Planck Society, and the Higher Education Funding Council for England. The SDSS Web Site is http://www.sdss.org/.
    The SDSS is managed by the Astrophysical Research Consortium for the Participating Institutions. The Participating Institutions are the American Museum of Natural History, Astrophysical Institute Potsdam, University of Basel, University of Cambridge, Case Western Reserve University, University of Chicago, Drexel University, Fermilab, the Institute for Advanced Study, the Japan Participation Group, Johns Hopkins University, the Joint Institute for Nuclear Astrophysics, the Kavli Institute for Particle Astrophysics and Cosmology, the Korean Scientist Group, the Chinese Academy of Sciences (LAMOST), Los Alamos National Laboratory, the Max-Planck-Institute for Astronomy (MPIA), the Max-Planck-Institute for Astrophysics (MPA), New Mexico State University, Ohio State University, University of Pittsburgh, University of Portsmouth, Princeton University, the United States Naval Observatory, and the University of Washington.
Funding for SDSS-III has been provided by the Alfred P. Sloan Foundation, the Participating Institutions, the National Science Foundation, and the U.S. Department of Energy Office of Science. The SDSS-III web site is http://www.sdss3.org/.
SDSS-III is managed by the Astrophysical Research Consortium for the Participating Institutions of the SDSS-III Collaboration including the University of Arizona, the Brazilian Participation Group, Brookhaven National Laboratory, Carnegie Mellon University, University of Florida, the French Participation Group, the German Participation Group, Harvard University, the Instituto de Astrofisica de Canarias, the Michigan State/Notre Dame/JINA Participation Group, Johns Hopkins University, Lawrence Berkeley National Laboratory, Max Planck Institute for Astrophysics, Max Planck Institute for Extraterrestrial Physics, New Mexico State University, New York University, Ohio State University, Pennsylvania State University, University of Portsmouth, Princeton University, the Spanish Participation Group, University of Tokyo, University of Utah, Vanderbilt University, University of Virginia, University of Washington, and Yale University.

\bsp

\label{lastpage}


\begin{thebibliography}{}

\bibitem[Abazajian et al.(2009)]{A09} Abazajian K. N. et al., 2009,
\apjs, 182, 543

\bibitem[Ade et al.(2014)]{Ade14} Planck collaboration: Ade P. A. R.,
Aghanim N., Armitage-Caplan C., Arnaud M. et al., 2014, \aap, 571, A16 

\bibitem[Ahn et al.(2014)]{A14} Ahn C. P. et al., 2014, \apjs, 211, 17 

\bibitem[Aller(1984)]{A84} Aller L. H., 1984, Physics of Thermal Gaseous
Nebulae. Dordrecht: Reidel

\bibitem[Alloin et al.(1979)]{A79} Alloin D., Collin-Souffrin S., Joly M., 
Vigroux L., 1979, \aap, 78, 200

\bibitem[Amor\'in et al.(2010)]{A10} Amor\'in R. O., P\'erez-Montero E.,
V\'lchez J. M., 2010, \apj, L128

\bibitem[Andrews \& Martini(2013)]{AM13} Andrews B. H., Martini P.,
2013, \apj, 765, 140

\bibitem[Baldwin et al.(1981)]{B81} Baldwin J. A., Phillips M. M., 
Terlevich R., 1981, \pasp, 93, 5


\bibitem[Bouwens et al.(2012)]{B12} Bouwens R. J., Illingworth G. D.,
Oesch P. A. et al., 2012, \apj, 754, 82


\bibitem[Cardamone et al.(2009)]{C09} Cardamone C. et al., 2009,
\mnras, 399, 1199

\bibitem[Cardelli et al.(1989)]{C89} Cardelli J. A., Clayton G. C.,
Mathis J. S.б 1989, \apj, 345, 245

\bibitem[Charlot \& Longhetti(2001)]{CL01} Charlot S., Longhetti M., 2001,
\mnras, 887

\bibitem[Coe et al.(2013)]{C13} Coe D., Zitrin A., Carrasco M. et al.,
2013, \apj, 762, 32




\bibitem[Cullen et al.(2014)]{C14} Cullen F., Cirasuolo M., McLure R. J.,
Dunlop J. S., Bowler R. A. A., 2014, \mnras, 440, 2300

\bibitem[Dawson et al.(2013)]{D13} Dawson K. S. et al., 2013, \aj, 145, 10 

\bibitem[Dopita et al.(2013)]{Do13} Dopita M. A., Sutherland R. S., Nicholls D. C.,
Kewley L. J., Fogt F. P. A., 2013, \apjs, 208, 10





\bibitem[Fioc \& Rocca-Volmerange(1997)]{FR97} Fioc M., 
Rocca-Volmerange B., 1997, \aap, 326, 950







\bibitem[Guseva et al.(2006)]{G06} Guseva N. G., Izotov Y. I., 
Thuan T. X., 2006, \apj, 644, 890

\bibitem[Guseva et al.(2007)]{G07} Guseva N. G., Izotov Y. I., 
Papaderos P., Fricke K. J., 2007, \aap, 464, 885

\bibitem[Guseva et al.(2009)]{G09} Guseva N. G., Papaderos P., 
Meyer H. T., Izotov Y. I., Fricke K. J., 2009, \aap, 505, 63

\bibitem[Hagen et al.(2014)]{H14} Hagen A. et al., 2014, \apj, 786, 59 

\bibitem[Heckman et al.(2005)]{H05} Heckman T. M. et al., 2005, \apj,
619, L35



\bibitem[Hoyos et al.(2005)]{Hoyos05} Hoyos C., Koo D. C., 
Phillips A. C., Willmer C. N. A., Guhathakurta P., 2005, \apj, 635, L21



\bibitem[Hunt et al.(2012)]{H12} Hunt L. K. et al., 2012, \mnras, 427, 906



\bibitem[Izotov et al.(1994)]{ITL94} Izotov Y. I., Thuan T. X.,
Lipovetsky V. A., 1994, \apj, 435, 647 

\bibitem[Izotov et al.(1997)]{ITL97} Izotov Y. I., Thuan T. X., \& 
Lipovetsky V. A., 1997, \apjs, 108, 1

\bibitem[Izotov et al.(2006)]{I06} Izotov Y. I., Stasi\'nska G., 
Meynet G., Guseva N. G., Thuan T. X., 2006, \aap, 448, 955

\bibitem[Izotov et al.(2011)]{I11a} Izotov Y. I., Guseva N. G.,
Thuan T. X., 2011, \apj, 728, 161



\bibitem[Izotov et al.(2014a)]{I14a} Izotov Y. I., Guseva N. G.,
Fricke K. J., Henkel C., 2014a, \aap, 561, A33

\bibitem[Izotov et al.(2014b)]{I14b} Izotov Y. I., Guseva N. G.,
Fricke K. J., Kr\"ugel E., Henkel C., 2014b, \aap, 570, A97

\bibitem[Jaskot \& Oey(2014)]{JO14} Jaskot A. E., Oey M. S., 2014,
\apj, 791, L19

\bibitem[Kakazu et al.(2007)]{Kakazu07} Kakazu Y., Cowie L. L., 
Hu E. M., 2007, \apj, 668, 853

\bibitem[Kauffmann et al.(2003)]{K03} Kauffmann G., Heckman T. M., 
Tremonti C. et al., 2003, \mnras, 346, 1055

\bibitem[Kennicutt(1998)]{K98} Kennicutt R. C., Jr., 1998, \araa, 36, 189

\bibitem[Kewley \& Dopita(2002)]{KD02} Kewley L. J., Dopita M. A.,
2002, \apjs, 142, 35

\bibitem[Kewley \& Ellison(2008)]{KE08} Kewley L. J., Ellison S. L.,
2008, \apj, 681, 1183

\bibitem[Kobulnicky \& Kewley(2004)]{KK04} Kobulnicky H. A., Kewley L. J.,
2004, \apj, 617, 240




\bibitem[Lara-L\'opez et al.(2010)]{L10} Lara-L\'opez M. A., Cepa J., 
Bongiovanni A. et al., 2010, \aap, 521, L53

\bibitem[Lee et al.(2006)]{L06} Lee H. et al., 2006, \apj, 647, 970


\bibitem[Leitherer et al.(1999)]{L99} Leitherer C. et al., 1999, \apjs, 123, 3


\bibitem[Maier et al.(2014)]{M14} Maier C., Lilly S. J., Ziegler B. L., 
P\'erez Montero E., Peng Y., Balestra I., 2014, \apj, 792, 3

\bibitem[Maiolino et al.(2008)]{M08} Maiolino R. et al., 2008, \aap, 488, 463

\bibitem[Manucci et al.(2010)]{M10} Manucci F., Cresci G., Maiolino R.,
Marconi A., Gnerucci A., 2010, \mnras, 408, 2115



\bibitem[Nicholls et al.(2013)]{N13} Nicholls D. C., Dopita M. A., Sutherland R.,
Kewley L. J., Palay E., 2013, \apjs, 207, 21

\bibitem[Oesch et al.(2014)]{O14} Oesch P. A., Bouwens R. J., 
Illingworth G. D. et al., 2014, \apj, 786, 108

\bibitem[Pagel et al.(1979)]{P79} Pagel B. E. J., Edmunds M. G., Blackwell D. E., Chun M. S., Smith G., 1979, \mnras, 189, 95

\bibitem[Palay et al.(2012)]{P12} Palay E., Nahar S. N., Pradhan A. K., Eissner W.,
2012, \mnras, 423, L35



\bibitem[Pettini \& Pagel(2004)]{PP04} Pettini M., Pagel B. E. J.,
2004, \mnras, 348, L59

\bibitem[Pettini et al.(2001)]{P01} Pettini M., Shapley A. E., 
Steidel C. C. et al., 2001, \apj, 554, 981

\bibitem[Pilyugin(2001)]{Pi01} Pilyugin L. S., 2001, \aap, 374, 412

\bibitem[Pilyugin \& Thuan(2005)]{PT05} Pilyugin L. S., Thuan T. X.,
2005, \apj, 631, 231

\bibitem[Pilyugin et al.(2010)]{P10} Pilyugin L. S., V\'ilchez J. M.,
Thuan T. X., 2010, \apj, 720, 1738

\bibitem[Refsdal et al.(1967)]{R67} Refsdal S., Stabell R., 
de Lange F. G., 1967, Mem. RAS, 71, 143

\bibitem[Salmon et al.(2015)]{Sal15} Salmon B., Papovich C., 
Finkelstein S. L. et al., 2015, \apj, 799, 183

\bibitem[Sanders et al.(2015)]{San15} Sanders R. L., Shapley A. E., Kriek M. et al., 2015, \apj, 799, 138




\bibitem[Smee et al.(2013)]{S13} Smee S. A. et al., 2013, \aj, 146, 32 


\bibitem[Stasi\'nska(2005)]{S05} Stasi\'nska G., 2005, \aap, 434, 507

\bibitem[Stasi\'nska(2006)]{St06} Stasi\'nska G., 2006, \aap, 454, L127



\bibitem[Steidel et al.(2014)]{S14} Steidel C. S. et al., 2014, \apj, 795, 165





\bibitem[Tremonti et al.(2004)]{T04} Tremonti C. et al., 2004, \apj, 613, 898

\bibitem[Troncoso et al.(2014)]{T14} Troncoso P. et al., 2014, \aap, 563, A58



\bibitem[York et al.(2000)]{Y00} York D. G. et al., 2000, \aj, 120, 1579

\bibitem[Zahid et al.(2011)]{Z11} Zahid H. J., Kewley L. J., Bresolin F., 2011, \apj, 730, 137

\bibitem[Zahid et al.(2012)]{Z12} Zahid H. J., Bresolin F., Kewley L. J., Coil A. L., 
Dav\'e R., 2012, \apj, 750, 120 

\bibitem[Zahid et al.(2013)]{Z13} Zahid H. J., Geller M. J., Kewley L. J., Hwang H. S.,
2013, \apj, 771, L19

\bibitem[Zahid et al.(2014a)]{Z14a} Zahid H. J. et al., 2014a, \apj, 791, 130

\bibitem[Zahid et al.(2014b)]{Z14b} Zahid H. J. et al., 2014b, \apj, 792, 75


\bibitem[Zhao et al.(2010)]{Zh10} Zhao Y., Gao Y., Gu Q., 2010, \apj, 710, 663


\end{thebibliography}
\end{document}